# Fresnel Number Concept and Revision of some Characteristics in the Linear Theory of Focused Acoustic Beams

Yu. N. Makov[1] and V.J. Sánchez-Morcillo[2]

[1] Department of Acoustics, Faculty of Physics, Moscow State University,

119899 Moscow, Russia.  Email: Yuri@makov.phys.msu.ru

[2] Departamento de Física Aplicada, Universidad Politécnica de Valencia, Carretera Nazaret-Oliva S/N, 46730 Grao de Gandia, Spain.  Email: victorsm@fis.upv.es

**ABSTRACT**

The advisability of the use of the Fresnel number as the measure (characteristic) of the ratio of diffraction and focusing properties for ultrasonic transducers and its radiated beams is proposed and demonstrated. Althought this characteristic is more habitual in optics, in acoustics the equivalent (mathematically although not fully in its physical meaning) parameter of linear gain is used as a rule. However, the preference and the more accuracy of the Fresnel number use is demonstrated here on the basis that the usual determination of the linear gain parameter ceases to correspond to the real value of the gain for low Fresnel number acoustic beams. It connects with the linear effect of axial maximum pressure shift from the geometrical focus towards the transducer. This effect is known for a long time, but here the analytical formulas describing this shift with a high accuracy for arbitrary Fresnel numbers are presented. As a consequence, also the analytical dependence of the real gain on the Fresnel number is obtained.

## 1. Introduction

In many respects the modern progress in acoustics, as well as in other scientific fields, is determined by the search of new nonlinear effects, and the investigation of the possible nonlinear generalization and transformation of the known linear effects [1, 2]. However, linear theories are often more than approximation in fact and can play an important role. The linear problems and conceptions are not only the starting points for their possible nonlinear generalizations, but also they constitute the fundamentals of every scientific discipline. Therefore, in cases of need the revision and correction of the known and traditional concepts and results in the linear theory can be of great importance. As an example of these arguments, there is the situation concerning the acoustic beam theory, where together with the successful development of the nonlinear theory [3, 4] some inaccurate conclusions and determinations are kept in the basic linear theory. Using the concept of Fresnel number these inaccuracies will be extracted and corrected here. The case in point is the inaccuracy of the known analytical expressions for the evaluation of the effect of on-axis maximum pressure shift from the geometrical focus towards the transducer and the discrepancy between the traditional determination of the linear gain parameter and its real value for low Fresnel number acoustic beams. The introduction of the Fresnel number concept (well known and widely used in optics) in conformity with the acoustic focused beams is presented in section 2. The improvement of the previously proposed expression for the axial maximum pressure shift, and the experimental verification of this result are presented in section 3. The revelation of the inaccuracy in the traditional determination of the linear gain parameter for low Fresnel number acoustic beams is given in section 4, where an approximate dependence of the real gain on the Fresnel number is obtained. Finally, some concluding remarks, which show the importance of the corrected results as the starting points for the study of nonlinear effects are discussed in section 5.

## 2. Fresnel number concept for acoustic focused transducers and its radiated beams.

The structure of the field radiated by acoustic focused sources in linear regime is determined, in many respects, by the ratio between the diffractional (Rayleigh) length $L_d = ka^2/2$ (where $a$ is the aperture radius and $k$ the wave number), which characterizes the degree of diffractional divergence, and the geometrical focal length (radius of curvature of the transducer) $R$, characterizing the focusing convergence. The quotient of these characteristics $L_d/R$ is the simplest non-dimensional parameter, which determines the relative but oppositely directed influences of these two effects. In acoustic beam theory this first direct and general sense of the given quotient is not fixed by the concrete value of the parameter, but traditionally it is used in other more specific and narrow interpretation, representing the value of the linear gain $G$ in the geometrical focal point of a focused transducer with not very large half-aperture angle $\alpha$ (with $\alpha$ smaller than 20º) under the Fresnel approximation. A more detailed analysis (see below) shows that, while keeping its first main meaning, the quotient $L_d/R$ do not account for the correct value of the real gain (determined as the ratio of on-axis maximum pressure to the pressure near the transducer surface) when $L_d \leq R$. This situation justifies the necessity to have an independent parameter for the estimation of the relative action of diffraction and focusing effects for focused acoustic beams. Such parameter is the Fresnel number, which is frequently used in the characterization of optical systems [5]. In the following, this parameter will be used for the characterization of acoustic beams and sources.

The Fresnel number $N_F$ is defined as the number of Fresnel zones fitted, when seen from a point of observation, into the bounded part of wave front determined by the aperture or screen on which the wave diffracts. In its turn, the Fresnel zones are the concentric annular regions on the wave front where each circular bound consists of points that are equidistant from the point of observation, and the paths from intersection points of common radius (curvilinear in general case) between neighbor circles on the surface of the wave front to the considered point of observation differ by half wavelength. In the simplest case of a plane circular region of the wave front, with radius $a$ and wavelength $\lambda$, the Fresnel number for a point on the axis of symmetry at a distance $z$ from the wave front is equal to the ratio of the square of the aperture radius to the square of the radius of the first Fresnel zone (the areas of all Fresnel zones are equal). Since the radius of the $m$-th Fresnel zone is given by $r_m = \sqrt{m\lambda z}$ [5], then

$$N_F(z) = \frac{a^2}{\lambda z} \qquad (1)$$

The concepts of Fresnel zones and Fresnel number give easily an estimate of the field in the point of observation as the result of the interference between waves arriving from different Fresnel zones.

For focused beams there is one preferred on-axis point, corresponding to the geometrical focus with coordinate $z = R$. For this point the concrete value of the Fresnel number is

$$N_F(z=R) \equiv N_F = \frac{a^2}{\lambda R} = \frac{L_d}{\pi R} \ . \tag{2}$$

This parameter, as the quotient of $L_d$ and $R$ (with the additional multiplier $\pi^{-1}$) characterizes the comparative influence of diffraction and focusing effects. Note, however, that in this focusing case the value of the Fresnel number does not allow to interpret outright the result of the interference field in the focal point. Since $N_F$ is the number of Fresnel zones fitted into the aperture plane AB of a concave focusing transducer (see Fig.1), viewed from the geometrical focus, then the phase difference between waves traveling from different points in AB is compensated by the corresponding phase difference of points at the concave wave front relative to the aperture plane AB.

For waves at the boundary of the transducer, the path difference from plane to concave surfaces is given by $AF - DF = AA'$. In this focusing case the Fresnel number $N_F$ can be interpreted as the relative (in half-wavelengths) "depth" $AA'$ of the transducer, which in the Fresnel approximation is given by

$$AA' = R - \sqrt{R^2 - a^2} \approx \frac{a^2}{2R} \tag{3}$$

and therefore

$$N_F = \frac{AA'}{\lambda/2} \ . \tag{4}$$

In this interpretation, the condition $N_F \gg 1$ corresponds to a strong effect of the initial wave front curvature (i.e. the effect of strong focusing) and the geometrical approach, which predicts the maximum of the field in the center of curvature $F$, where all rays converge, is justified. In the opposite case $N_F \ll 1$ the effect of the initial wave front curvature (i.e. the focusing effect) is insignificant, and the structure of the irradiated field approaches the result from a plane circular transducer.

Note that the separation of the focusing transducers (and the corresponding irradiated beams) into *high-* and *low-Fresnel-number* transducers (focused beams) is not equivalent, in the Fresnel approximation, to a large or small half-aperture angle $\alpha$ of the transducers, because in the expression for $N_F$ the certainly large value of $a/\lambda$ is multiplied by $\sin(\alpha) = a/R$, and for small angles this product may be large.

### 3. On-axis pressure maximum shift

The Fresnel number concept for a focused transducer provides an easy interpretation of the strong shift (from the geometrical focus towards the transducer) of the on−axis maximum pressure for the low-Fresnel-number transducers. Such focused transducers are close to plane transducers because of its small depth $AA'$ (defined in the previous section). On one side, it is known that, in the case of a plane circular transducer, the location $z_0$ of the first (the most far from transducer) on-axis field maximum is determined by the relation [3]

$$N_F(z_0) = \frac{a^2}{\lambda z_0} = 1. \tag{5}$$

On the other side, for a focused transducer in the low-Fresnel number limit, it holds

$$N_F = \frac{a^2}{\lambda R} \ll 1. \tag{6}$$

Then, comparing Eq. (5) and (6) it follows that $z_0 \ll R$, i.e. the location of the on-axis field maximum is much closer to the transducer than geometrical focus if $N_F \ll 1$.

Of course, the fact of the discrepancy (shift) between the geometrical focus and the main maximum pressure point is known in focused sound beam theory, but the case of strong shift for low-Fresnel-number transducers remains without due attention, and the analytical approximate expressions obtained for this shift in Refs. 6 and 7 underestimate the value of the shift in the indicated case. In order to show this, and with the aim of obtaining a more accurate expression for the shift of the maximum pressure, applicable to transducers with arbitrary $N_F$, we consider as a starting point of our analysis the exact analytical solution for the pressure complex amplitude $A(r,z)$, determined by the ordinary wave equation in the parabolic approximation (see, for example, Ref. 7)

$$A(r,z) = -\frac{ik}{z} \exp\left(\frac{ik}{2z} r^2\right) \int_0^a \exp\left(\frac{ik}{2z} r'^2\right) J_0\left(\frac{k}{z} rr'\right) A(r',0) r' dr', \qquad (7)$$

with the acoustic pressure being $p(r,z,t) = A(r,z) e^{ikz - i\omega t}$.

In the simplest case of constant pressure along the transducer, and parabolic phase accounting for the focusing effect, the initial condition reads

$$A(r',0) = p_0 \exp\left(-\frac{ikr^2}{2R}\right), \qquad (8)$$

for which the on-axis pressure distribution can be calculated from Eq. (7), and results

$$\left|\frac{p(0,\tilde{z})}{p_0}\right| = \left|\frac{2}{1-\tilde{z}} \sin\left(\frac{\pi N_F}{2} \frac{1-\tilde{z}}{\tilde{z}}\right)\right|, \qquad (9)$$

where the bars denote absolute value, $p_0$ is the constant pressure along the transducer aperture, and $\tilde{z} = z/R$ is a dimensionless coordinate along the beam axis. In this new coordinate, the geometrical focus is located at $\tilde{z} = 1$. Fig. 2 shows the on-axis pressure distribution as follows from Eq. (9), for a focusing transducer with $N_F = 1.28$ ($a = 1.5$ cm., $R = 11.7$ cm and work wavelength $\lambda = 0.15$ cm), which has been used in our experiments. Here the strong on-axis pressure maximum shift is visible. Taking the derivative of Eq. (9) and equating it to zero, the locations of the points of pressure extremum can be found as the solutions of the transcendental equation

$$\frac{\tan(X)}{X} = \frac{1}{\tilde{z}} \qquad (10)$$

where $X = \frac{\pi N_F}{2} \frac{1-\tilde{z}}{\tilde{z}}$.

The root of Eq. (10) for which the pressure distribution in Eq. (9) takes the largest value, corresponds to the location of the main maximum of on-axis pressure. The result of the numerical calculation of this root ($\tilde{z}_{max}$) from Eq. (10), as a function of the Fresnel number $N_F$, is shown with symbols in Fig. 3. For comparison, also the approximate analytical result

$\tilde{z}_{max} = 1 - 12/\pi^2 N_F^2$, given by Eq. (20) from Ref. 7, is shown (dashed line), which evidences the inaccuracy this expression for evaluating the shift in the low-Fresnel-number region.

Next, an improved approximate expression predicting the correct value of the shift in the whole range of $N_F$ is obtained. For this aim, it is convenient first to rewrite Eq. (10) in the form

$$\frac{\sin(X)}{X} = \left(\frac{2X}{\pi N_F} + 1\right) \cos(X). \tag{11}$$

The numerical solution of Eq. (10) shows that the values of $X(N_F, \tilde{z}_{max}(N_F))$ change from very small values for large $N_F$ until values close to unity for small $N_F$. On this basis, we can expand both sides of Eq. (11) into power series up to 3$^{rd}$ order in $X$. The operations with Eq. (11) for the expansion into series are preferable than with $\tan(X)$ in Eq. (10), because the alternate series for Eq. (11) converge better and faster. We will not perform here a special mathematical analysis for this convergence because we have the possibility to compare the results from the approximate expression and from the existing numerical solution.

After the expansion of Eq. (11) into series up to 3$^{rd}$ order in $X$ and the necessary reductions, the following quadratic equation for the evaluation of $\tilde{z}_{max}$ is obtained,

$$24\tilde{z}^2 = \pi^2 N_F^2 (3 - 4\tilde{z} + \tilde{z}^2), \tag{12}$$

whose solution (with the removal of the uncertainty) gives the dependence of the location $\tilde{z}_{max}$ of the on-axis maximum pressure with the Fresnel number $N_F$,

$$\tilde{z}_{max} = \frac{3\pi N_F}{2\pi N_F + \sqrt{\pi^2 N_F^2 + 72}} \tag{13}$$

Note that in the limit of large $N_F$ Eq. (13) reduces to $\tilde{z}_{max} \approx 1 - 12/\pi^2 N_F^2$, which is Eq. (20) from Ref. 7.

The obtained new approximate expression for the on-axis maximum pressure location $\tilde{z}_{max}$ is very accurate (very close to the numerical solution of Eq. (10)), as demonstrated in Fig. 3 (continuous line). The principal conclusion from the results in Fig. 3 is that for $N_F > 6$ (i.e., high-Fresnel-number focusing transducers) the main pressure maximum is very close to the geometrical focus, while for $N_F < 3$ (i.e. for low-Fresnel-number focusing transducers) the difference between these axial points is large, and the location of the main pressure maximum is strongly shifted towards the transducer. Note that, although the high-Fresnel-number focusing transducers are used more often in experimental and practical work, the low-Fresnel-number focusing transducers also have ordinary (not exotic) values of focal and aperture radius, and in fact they are serially manufactured. With one of them the strong on-axis pressure maximum shift has also been experimentally observed. For the above-mentioned focusing transducer, the on-axis pressure distribution was measured in the linear regime (Fig. 4). The linearity of the regime was controlled by the simultaneous inspection of the time profile, which was close to sinusoidal. The experimental result shows that, for the analyzed transducer with $N_F = 1.28$, the pressure maximum point locates at around 7 cm from the source, that for $R = 11.7$ cm results in a dimensionless coordinate for this point $\tilde{z}_{max} \approx 0.6$, which is close to theoretical value 0.67 (see Fig. 3).

## 4. Improved determination of the linear gain parameter for low Fresnel number acoustic beams.

It is well known that the traditional linear gain parameter in acoustic focused beam theory is defined as

$$G = L_d / R \qquad (13)$$

and indicates the degree of increase in the pressure at a characteristic point of the beam, with respect to the pressure near the transducer. From the normalized on-axis pressure distribution given in Eq. (9) it follows that the gain, as given in Eq. (13), is evaluated at the geometrical focal point. However, it is clear that the real gain $G_r$ must correspond to the maximum pressure point $\tilde{z}_{max}$, i.e.

$$G_r = \frac{p(0, \tilde{z}_{max})}{p_0}. \qquad (14)$$

For high-Fresnel-number focusing transducers, where the maximum pressure point is very close to the geometrical focal point, the difference between $G$ and $G_r$ is insignificant. However, for low-Fresnel-number focusing transducers this is not true, resulting in a considerable discrepancy between both values of the gain, as expected from the results shown in Fig. 2. Of course, the difference of the maximum value of the on-axis pressure distribution from the pressure in geometrical focus has been mentioned earlier (see e.g. [8]), but the determination of the linear gain (13) for the acoustical focused beams has remained without change.

The analytical dependence $G_r$ on $N_F$ is obtained after the substitution of $\tilde{z}_{\max}(N_F)$, given by Eq. (13), into Eq. (9),

$$G_r = \left| 2 \frac{(\pi N_F / 3) + \sqrt{2 + (\pi N_F / 6)^2}}{-(\pi N_F / 6) + \sqrt{2 + (\pi N_F / 6)^2}} \sin\left(-(\pi N_F / 6) + \sqrt{2 + (\pi N_F / 6)^2}\right) \right|. \quad (15)$$

This new and more accurate expression for the real gain $G_r(N_F)$, together with the simple linear dependence $G = \pi N_F$, are shown in Fig. 5, which evidences the incremental difference between $G$ and $G_r$ in the low Fresnel number region.

## 5. Conclusions

The correction, or even a more precise definition, of the existing scientific facts and results is important for the development of full and correct theories in all scientific fields, and in particular in acoustics. In this paper we present a revision of some linear effects for acoustic focused beams, which determines the correct results of the nonlinear development of these effects. Our experimental investigations show that, under increasing the driving voltage of the transducer, entering into the nonlinear regime of the acoustic beam, the on-axis pressure maximum point moves in the beginning (for low voltages) from its initial location towards the geometrical focal point, and then, with a further increase in the voltage, from the neighborhood of the geometrical focus, the pressure maximum moves backwards to the transducer. These results concerning the long excursion of the focal point at increasing driving voltages have been reported in [9], the initial stage of this nonlinear shift effect being determined by our corrected result from linear theory, presented in this article.

**Acknowledgements**

We wish to thank Juan Cruañes, Victor Espinosa and Jaime Ramis for the assistance in experimental investigation. This work was supported by the Russian Foundation for Basic

Researches (grant # 04-02-17009) and by the MEC of the Spanish Government, under the project FIS2005-07931-C03-02.## References

[1] Naugolnykh, K. A. and Ostrovsky, L., *Nonlinear Wave Processes In Acoustics*, Cambridge University Press (1998).

[2] Bakhvalov N.C., Zhileikin Ya.M. and Zabolotskaya E.A., *Nonlinear theory of sound beams*. American Institute of Physics, New York (1987)

[3] *Nonlinear Acoustics,* M. F. Hamilton and D. T. Blackstock (Ed). Academic, CA (1998). Chap. 8.

[4] O. V. Rudenko, O. A. Sapozhnikov, *Self-action effects for wave beams containing shock fronts*, Phys.-Uspekhi **174**, (2004) 973-989.

[5] M. Born and E. Wolf, *Principles of Optics*, Cambrige University Press (1999).

[6] H. T. O'Neil, *Theory of focusing radiators*, J. Acoust. Soc. Am. **21** (1949) 516-526.

[7] B. G. Lucas and T. G. Muir, *The field of a focusing source*, J. Acoust. Soc. Am. **72** (1982) 1289-1296

[8] M. V. Vinogradova, O. V. Rudenko, A. P Sukhorukov, *Wave Theory*, Moscow, "Nauka" (1990) (in Russian)

[9] Yu. Makov, V. Espinosa, V.J. Sánchez-Morcillo, J. Cruañes, J. Ramis and F. Camarena, *Strong on-axis focal shift and its nonlinear variation in low-Fresnel-number ultrasound beams*, J. Acoust. Soc. Am. (2006, in press)

FIGURE CAPTIONS

Figure 1. Geometrical scheme of a focused transducer and its basic parameters.

Figure 2. Theoretical on-axis pressure distribution in the linear regime of a focused transducer, obtained from Eq. (9), for $N_F = 1.28$.

Figure 3. Dependence of the position of the on-axis main pressure maximum on the Fresnel number. Symbols are the results from the numerical solutions of Eq. (10), continuous line corresponds to Eq. (4), and dashed line to the approximate result from Ref. (7). Axial distance in dimensionless units.

Figure 4. Experimental results of on-axis pressure distribution in the linear regime for the examined transducer.

Figure 5. Dependence of traditional and real gains with the Fresnel number.

Figure 1

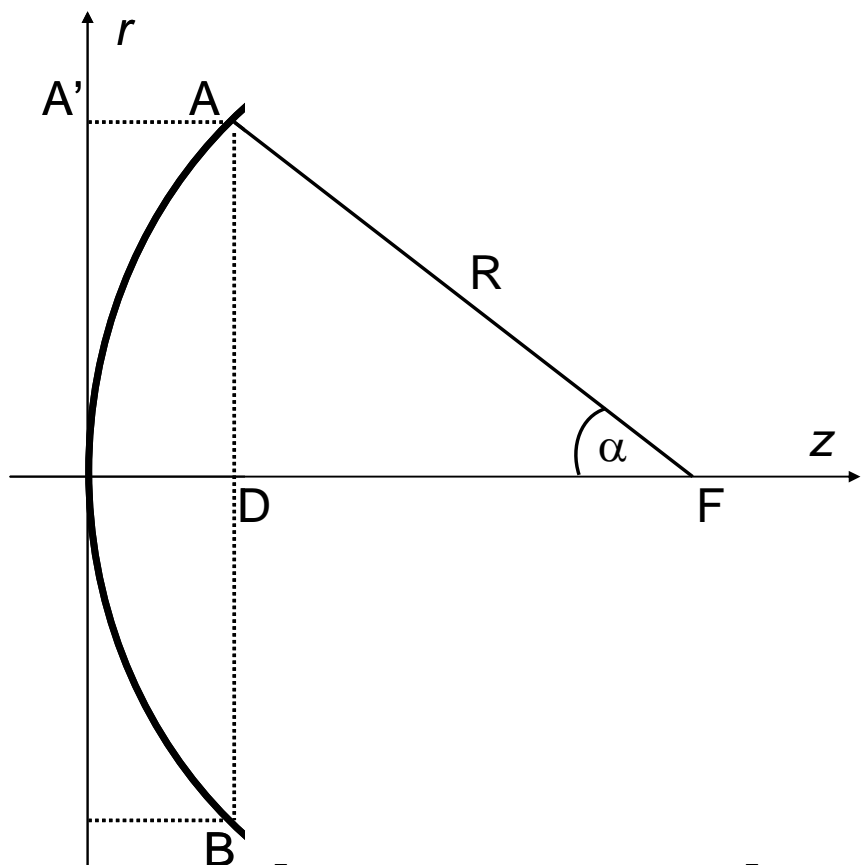

Figure 2

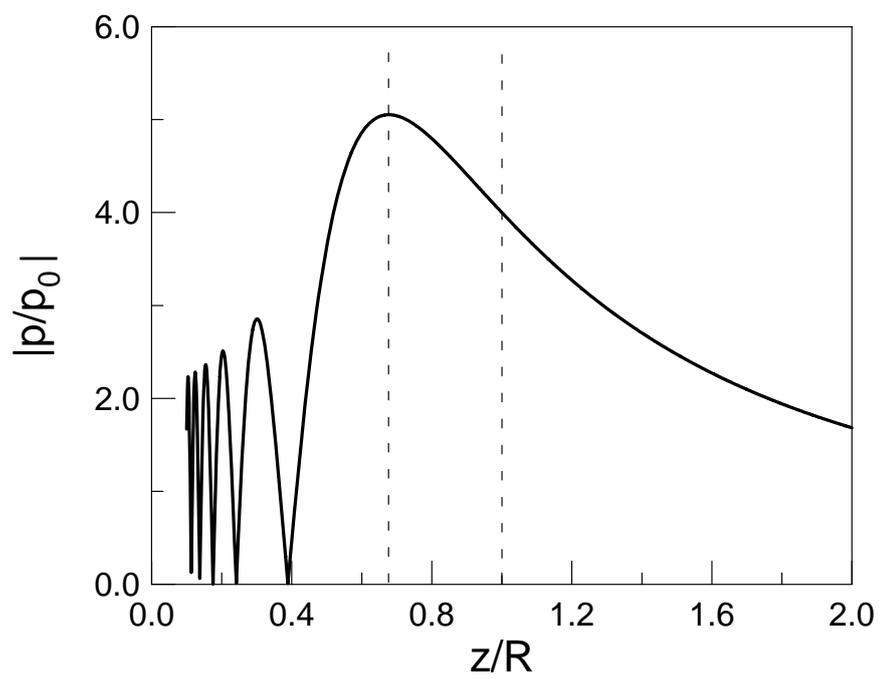

Figure 3

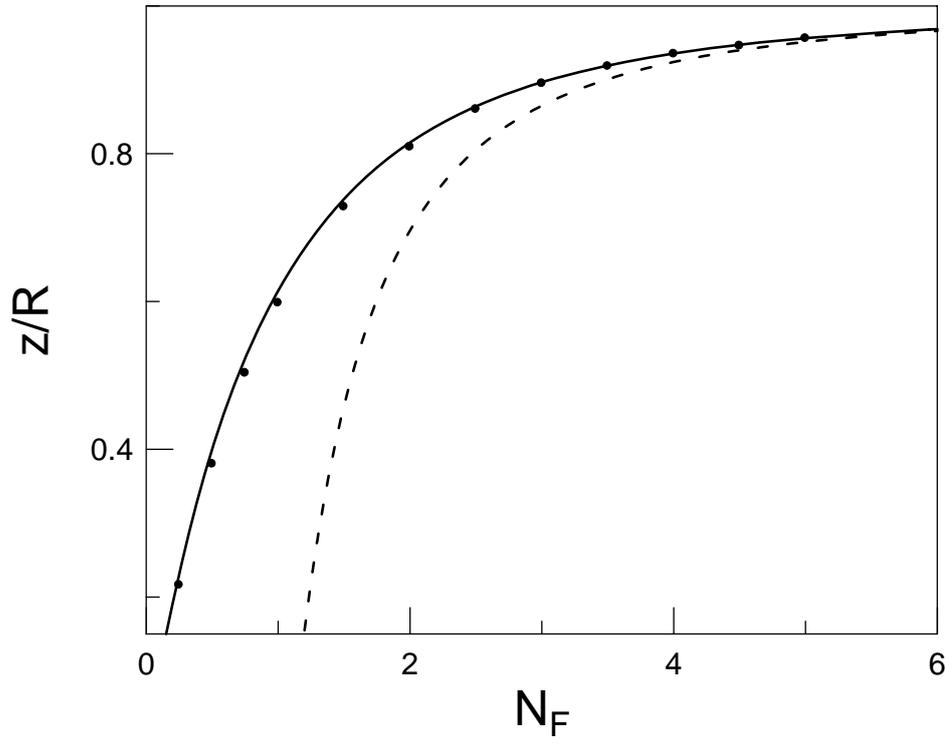

Figure 4

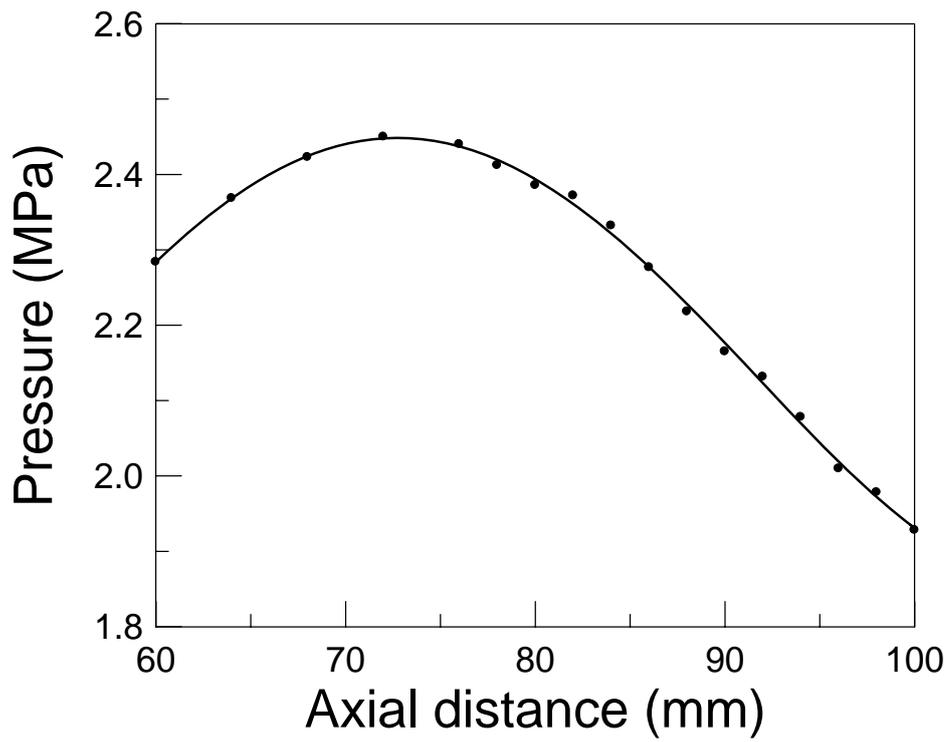

Figure 5

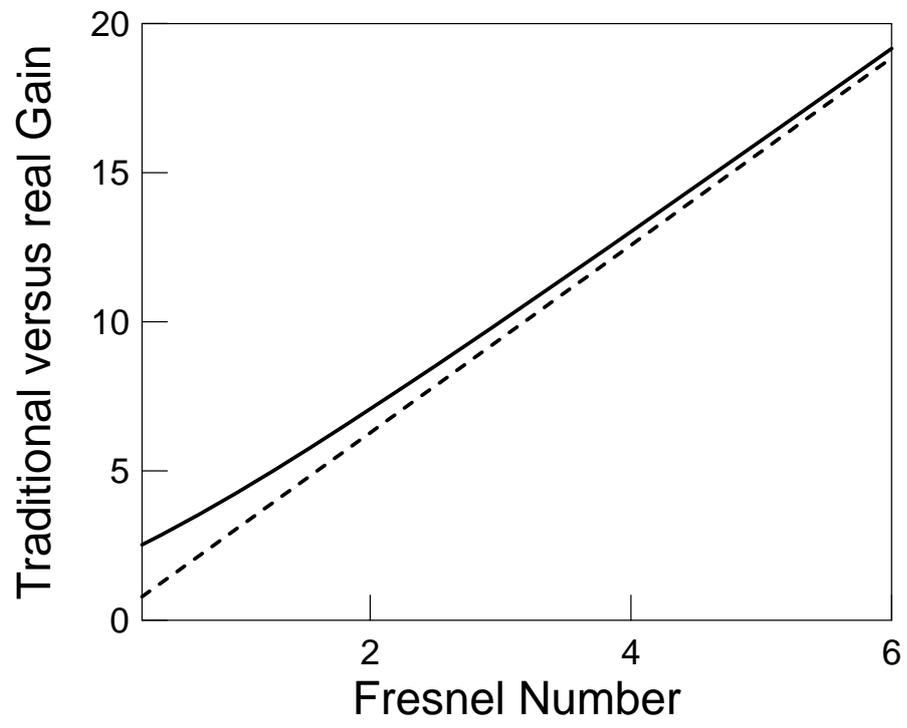